\documentclass[aip,jcp,reprint]{revtex4-2}

\usepackage{graphicx}
\usepackage{dcolumn}
\usepackage{bm}

\usepackage{amssymb,amsmath}
\usepackage{times}
\usepackage{subcaption}
\usepackage{hyperref}
\usepackage{booktabs}

\usepackage{color}

\usepackage[english]{babel}

\newcommand{\suptxt}[1]{$^{\text{#1}}$}
\newcommand{\subtxt}[1]{$_{\text{#1}}$}

\usepackage{caption}
\begin{document}


\title{Density-Matrix Embedding Based Multi-Configurational Perturbation Theory Approach to Single-Ion Magnets}

\author{Zhebin Guan}
\author{Hong Jiang}%
\thanks{To whom any correspondences should be addressed (Email: jianghchem@pku.edu.cn).}
\affiliation{Beijing National Laboratory for Molecular Sciences, College of Chemistry and Molecular Engineering, Peking University, Beijing 100871, China}

\date{\today}

\begin{abstract}
Multi-configurational wave-function theory (MC-WFT) that combines complete active space self-consistent field (CASSCF) approach with subsequent state interaction (SI) treatment of spin-orbit coupling (SOC), abbreviated as CASSCF-SO, plays important roles in microscopic understanding of single-ion magnets (SIMs) with different central transition metal or lanthanide ions and various coordination environments, but its application to SIMs with complex structure is severely limited due to its highly demanding computational cost. Density-matrix embedding theory (DMET) provides a systematic and mathematically rigorous framework to combine low-level mean field approaches like Hartree-Fock and high-level MC-WFT methods like CASSCF-SO, which is particularly promising to SIMs. As a continuation of our previous work on DMET+CASSCF for $3d$ SIMs (Ai, Sun, and Jiang, \textit{J. Phys. Chem. Lett.} 2022, 13, 10627), we extend the methodology by considering dynamic correlation on top of CASSCF using the second-order $n$-electron valence perturbation theory (NEVPT2) in the DMET framework, abbreviated as DMET+NEVPT2, and benchmark the accuracy of this approach to molecular magnetic anisotropy in a set of typical transition metal complexes. We found that DMET+NEVPT2 can give the results very close to all-electron treatment, and can be systematically improved for higher accuracy by expanding the region treated as the central cluster, while the computation cost is dramatically reduced due to the reduction of the number of orbitals by DMET construction. Our findings suggest that DMET is capable of accounting for most of the dynamic correlation that is important for magnetic anisotropy in typical SIMs, and can be useful for further high-accuracy spin-phonon study and high-throughput computations.

\end{abstract}

\maketitle


\section{\label{sec:introduction}Introduction}

Single-molecule magnets (SMMs), which exhibit magnetic hysteresis at the molecular level \cite{Benelli2015}, hold immense promise for next-generation electronic and spintronic technologies, with potential applications in information storage, quantum sensing, quantum computing, and so on \cite{Moreno-Pineda2021}. A key feature of SMMs is molecular magnetic anisotropy, often characterized by zero-field splitting (ZFS) parameters, which describes the directional dependence of a molecule's response to an external magnetic field and its ability to retain magnetization. For practical device applications, achieving large magnetic anisotropy is critical, as it ensures the persistence of finite magnetization at higher temperatures—a primary objective since the inception of SMMs \cite{Rinehart2011, Sarkar2020, Raza2023, WangC2023}. Early strategies focused on coupling magnetic centers to create large magnetic moments through polynuclear transition metal complexes, which, however, has proven to be inefficient, as the anisotropy contributions from different magnetic ions within the molecule tend to cancel out each other such that the total magnetic anisotropy does not scale with the total spin \cite{Waldmann2007, Neese2011}. The discovery of single-ion magnets (SIMs) \cite{Ishikawa2003}, where molecular anisotropy arises from a single magnetic center, marked a significant breakthrough. This design allows for easier control of anisotropy through crystal field tuning, leading to the synthesis of numerous SIMs based on transition metal or lanthanide ions \cite{WangY2024,FengM2018}. As a result, records for the effective spin-reversal barrier (\(U_{\rm eff}\)) and blocking temperature (\(T_{\rm B}\)) have been continually surpassed \cite{Goodwin2017,Guo2018}. While most recent efforts have focused on lanthanide SIMs, especially those based on Dy(III) \cite{WangY2024}, there are also continuous interest in \(3d\)-transition metal-based SIMs as well \cite{FengM2018}, as exemplified by recent discovery of a linear cobalt (II) complex with intreguing slow magnetic relaxation \cite{Bunting2018}. A detailed microscopic understanding of \(3d\)-SIMs, including their electronic structure and spin-phonon relaxation mechanisms \cite{Lunghi2022, Chibotaru2023}, is therefore essential for further progress.

Quantum chemistry calculation based on multi-configurational wave-function theory (MC-WFT) has played a pivotal role in advancing fundamental research on SMMs by providing atomic-level insights into their electronic structure, magnetic properties, and spin dynamics, which can not only elucidate the mechanisms of magnetic relaxation and quantum tunneling but also guide the synthesis of novel molecules with tailored properties \cite{Atanasov2015, Ungur2015, Lunghi2022, Chibotaru2023}. However, significant challenges remain in the first-principles calculation of SMMs. Spin-orbit coupling (SOC) is the fundamental origin of magnetic anisotropy in SMMs and must be accurately incorporated into first-principles calculations. In methods such as state-interaction spin-orbit coupling (SISO) for the consideration of SOC \cite{Malmqvist2002, Atanasov2015}, highly accurate wavefunctions and energies for tens or hundreds of low-lying spin-free electronic states are needed to obtain matrix elements of the full Hamiltonian including the SOC term, often requiring computationally intensive MC-WFT approaches such as state-averaged (SA) complete active space self-consistent field (CASSCF) with subsequent consideration of dynamic correlation by multi-reference perturbation theory (MRPT). Such a composite MC-WFT approach, has been widely used in theoretical study of molecular magnetism \cite{Atanasov2015, Chibotaru2023}, but its application to systems with complex molecular structures with hundreds or thousands of atoms is still very challenging, and becomes infeasible to simulate directly SMMs in the crystal phase or deposited on surfaces that are crucial to practical device implementation \cite{Cinchetti2017}. Rational design of novel SIMs requires exploring the vast chemical space of SIMs with diverse coordination environments and ligand structures \cite{Mariano2024}, which calls for developing efficient and accurate computational methods with the accuracy of MC-WFT.

A highly promising approach to addressing these challenges is the use of quantum embedding methods in quantum chemistry \cite{Sun2016,Jones2020}. The essence of quantum embedding theory is to partition the whole system into a cluster of interest (often termed as ``impurity'') and its surrounding environment, denoted as $\mathcal{I}$ and $\mathcal{E}$, respectively. While the entire system is treated with lower-level methods like Hartree-Fock (HF) or Kohn-Sham (KS) density-functional theory (DFT), high-level MC-WFT methods can be applied to the embedded impurity subsystem, which, with a proper treatment of impurity-environment interactions by a certain embedding technique, can give an accurate description of properties of the whole system. Among various embedding techniques that have been developed, including those based on electron density \cite{Jacob2014,Wesolowski2015}, density matrices \cite{Knizia2013, Fornace2015, Welborn2016, Yu2017}, and Green's functions \cite{Kotliar2006, Zgid2017,Ma2021}, density-matrix embedding theory (DMET) proposed by Chan and coworkers \cite{Knizia2012,Knizia2013,Wouters2016} is particularly attractive, as it can be formulated in a mathematically rigorous manner and provides a second-quantized Hamiltonian for the embedded subsystem, enabling seamless integration with high-level quantum chemistry solvers. In the past decade, DMET has been actively pursued by several groups and demonstrated excellent performance across diverse chemical systems \cite{Bulik2014, Wouters2016,Pham2018,Mitra2022,Cui2020,Cui2022,Haldar2023, Ai2022,CaoC2023}.

For SIMs, the effectiveness of the spin Hamiltonian \cite{Neese2002, Atanasov2015, Chibotaru2013} suggests that magnetic properties of these systems are primarily governed by the central metal atom and its interactions with surrounding ligands. This provides a natural partitioning scheme, where the transition metal center is treated as the impurity and the ligands as the environment. Using this strategy, we developed in our previous work \cite{Ai2022} a DMET-based quantum embedding approach to SIMs that uses restricted open-shell Hartree-Fock (ROHF) as the low-level solver and CASSCF-SO as the high level-solver, denoted as DMET+CASSCF, which was found to accurately describe ZFS parameters of Co\(^{\text{II}}\)- and Fe\(^{\text{II}}\)-based SIMs with an error typically below 3 cm\(^{-1}\) when compared to all-electron CASSCF-SO results. It should be mentioned that other embedding methods, such as auxiliary-field quantum Monte Carlo (AFQMC) with distance-based embedding \cite{Eskridge2023} and coupled-cluster theory (CCSD) embedded in density-functional theory (DFT) \cite{Alessio2024}, have also been applied to SIMs. 

In this work, we extend the DMET+CASSCF approach by considering dynamic correlation on top of CASSCF using the second-order $n$-electron valence perturbation theory (NEVPT2) \cite{Angeli2001a} in the DMET embedded impurity subspace, and benchmark the accuracy of this approach, abbreviated as DMET+NEVPT2, to molecular magnetic anisotropy in a set of typical transition metal complexes. The article is structured as follows: Section \ref{sec:method} introduces the quantum chemical methods and the DMET scheme used in this study. Section \ref{sec:results} presents our benchmark results for DMET calculations on systems with different central metal atoms and coordination environments, along with an analysis of the sources of error in the DMET approach. Finally, Section \ref{sec:summary} summarizes main findings of this work and gives some general remarks on future directions for further development.

\section{\label{sec:method}Theoretical Methods and Computational Details}

\subsection{Multi-configurational quantum chemistry approach to $3d$ SIMs}

In single-molecule magnets, a few lowest lying electronic states account for the magnetic properties of the molecule, and therefore it is possible to use an spin Hamiltonian to capture energy levels of those states and interaction among them, which can be used to interpret experimental observations \cite{Atanasov2015, Chilton2022, Chibotaru2023}. In $3d$ single-ion magnets, since the crystal-field splitting of electronic states is often significantly larger than the spin-orbit coupling \cite{Abragam1970}, the lowest states can be understood as the SOC-split multiplet of the electronic ground state, and form a subspace that have the same dimension as the ground state spin multiplicity. It is then convenient to use the spin Hamiltonian of the following form to describe magnetic properties \cite{Neese2002,Atanasov2015, Chibotaru2023}
\begin{equation}
    \hat{H}^{\rm{ZFS}} = \hat{\bm{S}} \cdot \mathbf{D} \cdot \hat{\bm{S}},
\end{equation}
where $\hat{\bm{S}}$ is the spin operator and $\mathbf{D}$ is a tensor matrix accounting for the zero-field magnetic anisotropy. Since $\mathbf{D}$ is real and symmetric, it is possible to choose a proper molecular coordinate system such that $\mathbf{D}$ is diagonal \cite{Graaf2016}. By a re-defition of energy zero, $\mathbf{D}$ can be taken in a traceless form, leading to a simplified spin Hamiltonian in the following form \cite{Atanasov2015}
\begin{equation}
    \hat{H}^{\rm{ZFS}} = D(\hat{S}_z^2 - \frac{1}{3}\hat{S}^2) + E(\hat{S}_x^2 + \hat{S}_y^2),
\end{equation}
where $D$ and $E$ are known as axial and rhombic anisotropy parameter, respectively, also known as zero-field splitting (ZFS) parameters. $D$ mainly determines the energy splitting between the maximally and minimally polarized states through $\Delta E = DS^2$ or $\Delta E = D(S^2-\frac{1}{4})$ for integer and half integer $S$, respectively, and $E$ indicate the mixing between different $M_S$ states in the eigenstates. To exhibit SMM behaviors, $D$ should be negatively large to stabilize the spin polarized state, and $E$ be as small as possible to reduce quantum tunneling of magnetism \cite{Atanasov2015}.


To describe SMMs in \emph{ab-initio} quantum chemistry calculation, it is necessary to treat SOC accurately. In this work, the SOC is taken into account within the widely used two-step scheme \cite{Atanasov2015}, also termed as state interaction spin-orbit (SISO) coupling method \cite{Malmqvist2002, Sayfutyarova2016}. One first solves Schr{\"o}dinger equation of a scalar relativistic many-electron Hamiltonian within a certain approximation, which leads to a set of spin-pure (or pre-SOC) wavefunctions with well defined spin quantum numbers ($S$ and $M_S$), $|\Psi_{I}^{SM}\rangle$ and corresponding energies $E^S_I$, with $I$ denoting the extra index needed. The latter are then used as the many-electron basis to expand SOC wavefunctions
\begin{equation}
 |\bar{\Psi}_K \rangle = \sum_{I, S, M} C_{ISM} |\Psi_{I}^{SM}\rangle,
\end{equation}
and the expansion coefficients are obtained by diagonalizing the matrix representation of the full Hamiltonian with the SOC term, i.e. the so-called ``master matrix''\cite{Atanasov2015}
\begin{equation}
    H_{I'S'M', ISM} = E_I^S \delta_{II'} \delta_{SS'} \delta_{MM'} +  \langle \Psi_{I'}^{S'M'} | \hat{H}_{\rm{SOC}} | \Psi_{I}^{SM} \rangle.
    \label{eq:master-matrix}
\end{equation}

In this work, we use the effective Hamiltonian theory approach \cite{Maurice2009} to extract the spin Hamiltonian parameters from \textit{ab initio} SOC wavefunctions. The lowest $2S+1$ SOC states $\{|\bar{\Psi}_K\rangle, K=1,2,\cdots,2S+1\}$ are projected onto the ground state multiplet space $\{|\Psi_0^{S M}\rangle, M= -S, -S+1, \cdots, S\}$, and further orthonormalized to $\left\{ |\tilde{\Psi}_K \rangle\right\}$, which are used define the effective Hamiltonian
\begin{equation}
    \hat{H}^{\rm{eff}} = \sum_K^{2S+1} |\tilde{\Psi}_K \rangle \bar{E}_K \langle \tilde{\Psi}_K|
\end{equation}
The parameters $D$ and $E$ can then be extracted by comparing the matrix elements of $\hat{H}^{\rm{eff}}$ and $\hat{H}^{\rm{ZFS}}$. We note that there are other more sophisticated methods available to build the spin Hamiltonian based on \textit{ab initio} calculation \cite{Chibotaru2012,Atanasov2015, Malrieu2014}. Since the main focus of this work is the assessment of the DMET accuracy, the findings presented below will not be affected by the particular method used to extract ZFS parameters.

In the SISO approach to SIMs, all states that are expected to have significant contribution to magnetic anisotropy should be included for the construction of the master matrix (eq. \ref{eq:master-matrix}). Those states are naturally included in the complete active space spanned by $3d$ orbitals of the central transition metal ion, in which all possible occupations of $3d$ electrons within the chosen set of active orbitals (their number denoted as $n_{\rm act}$) are taken into consideration. This picture lead to the use of CASSCF as the most widely used method \cite{Atanasov2015} to generate pre-SOC states for further state interaction treatment of SOC.
To further consider dynamic correlation with inactive (doubly occupied) and virtual orbitals in CASSCF, whose numbers are denoted as $n_{\rm inact}$ and $n_{\rm vir}$, respectively, we use the strongly contracted second-order $n$-electron valence perturbation theory(NEVPT2) \cite{Angeli2001a} in the quasi-degenerate formulation \cite{Angeli2004} to calculate energy corrections to all pre-SOC states.
The computational scaling for NEVPT2 comes from several aspects. In calculating the energies of perturbation  wavefunctions, an auxiliary matrix depending on four-particle density matrix is used, leading to $O(n_{\rm act}^8)$ scaling \cite{Angeli2002}. However, this scaling is not the major concern for $3d$-SIMs, since often the minimal active space of five $3d$ orbitals is used. The more demanding scaling is the number of perturbation wavefunctions that scales as $O(n_{\rm inact}^2 n_{\rm vir}^2)$ \cite{Mitra2021}. In the SISO consideration of SOC, one needs to calculate the energy correction to tens or hundreds of pre-SOC states, which gives an additional prefactor of the same magnitude. As a result, the computational cost for SIMs at the NEVPT2 level can be quite heavy, especially for large systems with hundreds of atoms. The main goal of this work is to reduce the factor related to the number of perturbation wavefunctions with little loss of accuracy by using the DMET technique, as will be discussed in later sections.

\subsection{Density matrix embedding theory based CAS approaches to SIMs}

In $3d$ SIMs, strongly correlated orbitals are localized on the central transition metal (TM) ion, and the lowest energy excitations are related to rearrangement of $3d$ electrons. However, the interactions between the ligand and the TM ion should not be simply treated as electrostatic ones, since metal-ligand covalent interaction is important for energy levels of transition metal complexes \cite{Singh2017}, and correlations between electrons in the TM center and those occupying ligand orbitals should be adequately treated to obtain quantitatively accurate results. This forms a well-defined localized strong correlation problem for which a quantum embedding treatment like DMET \cite{Sun2016} is particularly promising.

In the DMET+CASSCF approach to SIMs proposed in our previous work \cite{Ai2022}, we choose orthonormalized local orbitals (LOs) centered on the TM center as the impurity ($\mathcal{I}$), and all other orbitals centered on the ligand groups as the environment ($\mathcal{E}$), with their numbers denoted as $n_{\rm I}$ and $n_{\rm E}$, respectively. It should be noted that there are different variants of LOs that can be used for such partition (see, e.g. Ref. \citenum{Cui2020}), and we use the  L{\"o}wdin orthogonalized atomic orbitals, since they can be obtained in a most straightforward manner without involving sophisticated localization treatment. Based on the Schmidt decomposition of the mean-field ground state Slater determinant obtained \cite{Knizia2012} from restricted open-shell Hartree-Fock (ROHF), we can extract a set of orbitals from $\mathcal{E}$ that entangle strongly with $\mathcal{I}$, termed as ``bath'' orbitals, as well as a set of unentangled doubly occupied (so-called ``core'') orbitals in $\mathcal{E}$.
Bath orbitals can be derived from the ROHF ground state wave function $\Phi_0^{\rm ROHF}$ in several different ways that are mathematically equivalent \cite{Wouters2016}. In this work, we obtain bath orbitals by diagonalizing the environment part of the one-particle reduced density matrix corresponding to $\Phi_0^{\rm ROHF}$ \cite{Wouters2016}
\begin{equation}
   D^\mathcal{E}_{\mu\nu} =  \left \langle \Phi_0^{\rm ROHF} \right| \sum_\sigma \hat{c}_{\nu \sigma}^\dagger \hat{c}_{\mu \sigma} \left| \Phi_0^{\rm ROHF} \right \rangle, \quad (\mu, \nu \in \mathcal{E}),
\end{equation}
\begin{equation}
   \mathbf{D}^\mathcal{E} \mathbf{c}_k  = \zeta_k \mathbf{c}_k,
\end{equation}
which yields three sets of orbitals in $\mathcal{E}$:
\begin{enumerate}
    \item orbitals with eigenvalues $\zeta_k = 2$: the environment orbitals that are fully occupied, i.e. core orbitals;
    \item orbitals with eigenvalues $\zeta_k = 0$: the environment orbitals that are always unoccupied;
    \item orbitals with eigenvalues $0< \zeta_k < 2$: the environment orbitals that are fractionally occupied.
\end{enumerate}
Since fractional occupation indicates the entanglement between $\mathcal{I}$ and $\mathcal{E}$ \cite{Wouters2016}, the third set of orbitals are then identified as bath orbitals, denoted as $\mathcal{B}$. The first two sets of orbitals, denoted as $\mathcal{E}_{\rm core}$ and  $\mathcal{E}_{\rm vir}$, respectively, are un-entangled occupied and virtual orbitals purely residing in the environmental region. One can then define the embedded impurity space $\mathcal{I}_{\rm emb} \equiv \mathcal{I}\oplus \mathcal{B}$. Under the condition $n_{\rm I} < n_{\rm occ} < n_{\rm E}$, with $n_{\rm occ}$ denoting the number of occupied orbitals, one can prove \cite{Wouters2016} that there are at most $n_{\rm I} + n_{\rm o}$ bath orbitals (with $n_{\rm o}$ being the numbers of open-shell orbitals). Therefore the total number of orbitals in $\mathcal{I}_{\rm emb}$ to be treated in high-level calculation like CASSCF-SO is always less than $2n_{\rm I}+n_{\rm o}$, and does not grow with the size of environment. In practice, we use a small finite criterion $\epsilon$ and identify orbitals with eigenvalues falling in the range of $(\epsilon, 2-\epsilon)$ as bath orbitals. In this work, $\epsilon=10^{-13}$ is used. 

We can then obtain the embedded impurity Hamiltonian by projecting the total Hamiltonian to the embedded impurity subspace $\mathcal{I}_{\rm emb}$ \cite{Wouters2016}, which is represented in the second quantization form as
\begin{equation}
\begin{aligned}
    \hat{H}_{\rm emb} & = \sum_{i,j \in \mathcal{I}_{\rm emb}} \sum_\sigma h_{ij} \hat{c}_{i\sigma}^\dagger \hat{c}_{j \sigma} \\
    & + \frac{1}{2} \sum_{i,j,k,l\in \mathcal{I}_{\rm emb}} \sum_{\sigma\sigma'} \langle ij|kl\rangle \hat{c}_{i\sigma}^\dagger \hat{c}_{j\sigma'}^\dagger \hat{c}_{l\sigma'} \hat{c}_{k\sigma}.
\end{aligned}
\end{equation}
Here $\langle ij|kl\rangle$ are two-electron Coulomb integrals, and $h_{ij}$ are matrix elements of the single-particle operator
\begin{equation}
    \hat{h} = -\frac{1}{2} \nabla^2 - \sum_{I \in \mathcal{I+E}} \frac{Z_I}{|r - R_I|} + \sum_{a\in \mathcal{E}_{\rm core}} (2\hat{J}_a - \hat{K}_a),
\end{equation}
with $\hat{J}_a$ and $\hat{K}_a$ being Coulomb and exchange operators of frozen environmental core orbitals that accounts for  mean-field Coulomb and exchange interactions between electrons in $\mathcal{I}_{\rm emb}$ and those occupying core orbitals, i.e. $\langle i |\hat{J}_a | j\rangle = \langle ia| ja\rangle$ and $\langle i |\hat{K}_a | j\rangle = \langle ia| aj\rangle$. It is worthwhile to note that up to a constant energy shift, $\hat{H}_{\rm emb}$ has essentially the same form as the Dyall Hamiltonian \cite{Dyall1995}, which indicates some conceptual link between DMET and the CAS problem. But instead of treating all orbitals in $\mathcal{I}_{\rm emb}$ as active orbitals, in the DMET+NEVPT2 approach to $3d$-SIMs, we choose from the ROHF solution of $\hat{H}_{\rm emb}$ a small set of orbitals with $3d$ characters as the active orbitals, and conduct the NEVPT2-SISO calculation fully within $\mathcal{I}_{\rm emb}$.

\begin{figure}
    \centering
	\includegraphics[width=0.48\textwidth]{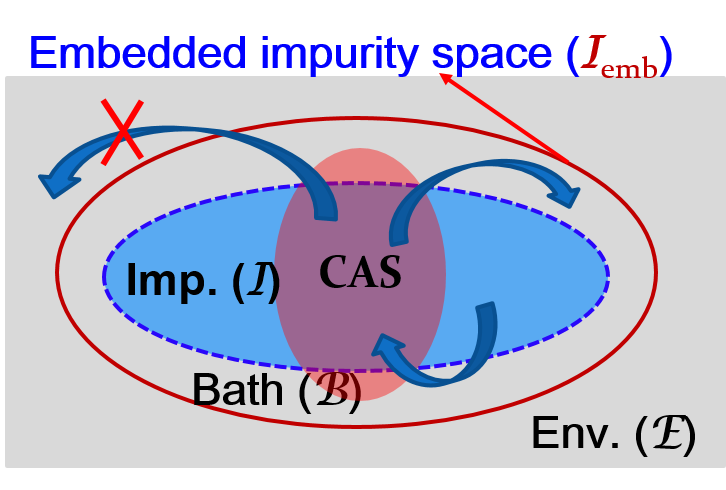}
	\caption{Schematic illustration of the physical essence underlying the DMET+NEVPT2 approach. The blue dashed ellipse indicates a primary partition of the whole system into the impurity ($\mathcal{I}$) and environment ($\mathcal{E}$) and the red solid one indicates the embedded impurity space ($\mathcal{I}_{\rm emb}$). The complete active space (CAS) considered in CASSCF is spanned by a small selected set of orbitals within $\mathcal{I}_{\rm emb}$.}
    \label{fig:DMET_orbitals}
\end{figure}

The physical essence underlying the DMET+NEVPT2 approach is illustrated in Figure \ref{fig:DMET_orbitals}. Using DMET amounts to restricting the orbitals involved in CASSCF orbital rotation and NEVPT2 perturbation wavefunction construction \cite{Pham2018}. The dynamical correlations within $\mathcal{I}_{\rm emb}$ is taken into account, but those involving unentangled occupied (core) and virtual environmental orbitals are neglected. Obviously the efficacy of such treatment critically depends on the quality of $\mathcal{I}_{\rm emb}$. In our previous work \cite{Ai2022}, we found that ROHF using conventional self-consistent field (SCF) techniques like direct inversion of iterative subspace (DIIS) \cite{Pulay1982} often converges to meta-stable solutions with significantly delocalized spin polarization in the ligand region of TM complexes, and usually results in poor performances in the DMET treatment, which, however, can be effectively overcome by using a new SCF scheme called regularized(R)-DIIS (see Ref.\citenum{Ai2025} for more details). Using R-DIIS can ensure to obtain the ROHF solution with both a lower energy and spin-polarization strongly localized on the transition metal ion for all systems considered in this work. One can also improve the accuracy of the DMET treatment systematically by expanding the size of the primary impurity space $\mathcal{I}$, which, for $3d$-SIMs in particular, means including localized atomic orbitals centered on ligand atoms directing bonded to the TM ion in $\mathcal{I}$.

\begin{figure*}[!htb]
    \centering
    \includegraphics[width=0.95\textwidth]{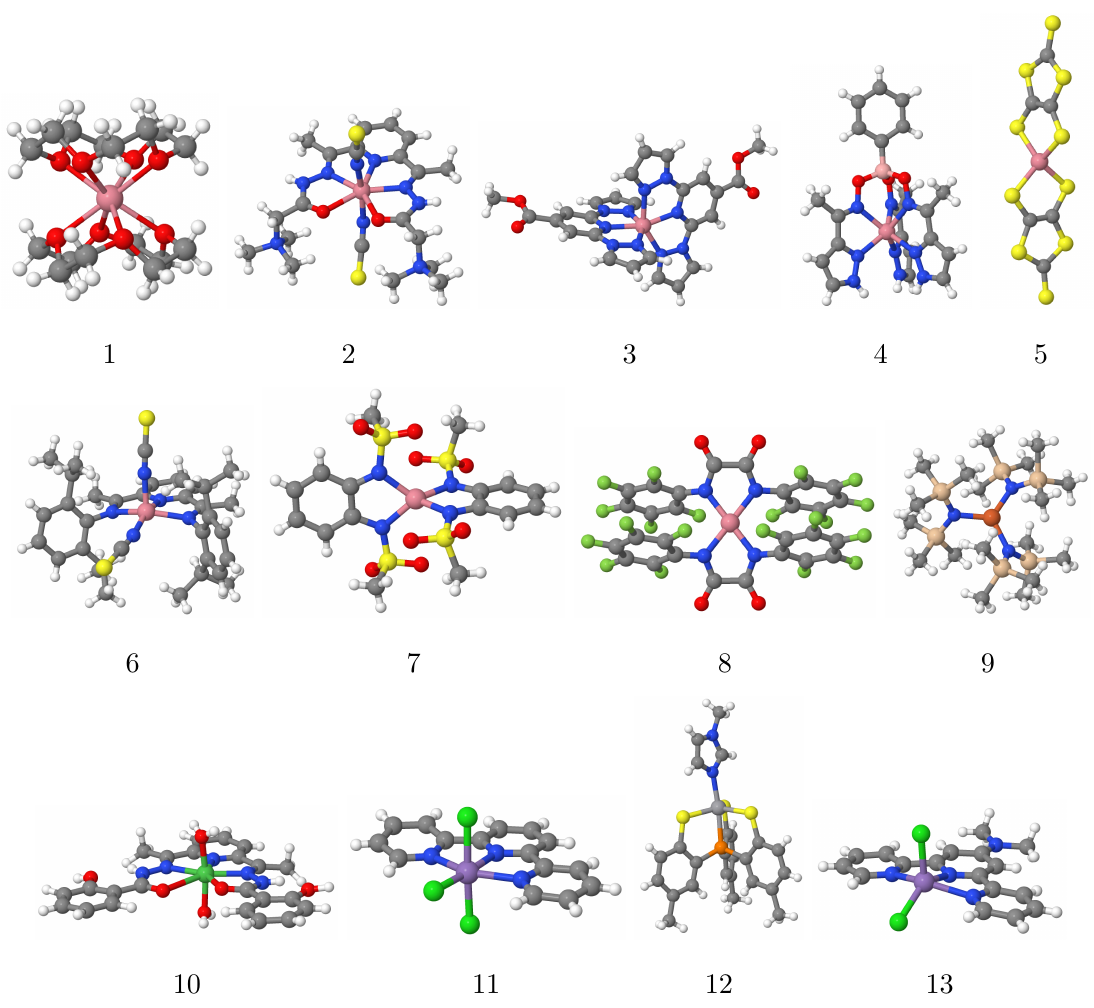}
	\caption{Illustration of molecular structures of transition metal complexes taken from X-ray diffraction measurements \cite{Chen2014, Darmanovic2019, Rigamonti2018, Novikov2015, Fataftah2014, Jurca2011, Rechkemmer2016, Gupta2023, Eichhofer2014, Ruamps2013, Duboc2010, Ye2010, Zein2008}.
	Color codes: pink for Co, brown for Fe, dark green for Ni, purple for Mn, orange for V, red for O, blue for N, yellow for S, light pink for B, pale green for F, light green for Cl, light brown for Si, gray for C, white for H.}
	\label{fig:struct}
\end{figure*}

\subsection{Computational details}
We have selected a set of typical transition metal complexes with different coordination environments to benchmark the accuracy of combining DMET with CASSCF and NEVPT2 methods. The main focus is on  Co\suptxt{II} complexes, which perform best among $3d$ SIMs, while a few other complexes with Ni, Fe, Mn and V that exhibit magnetic anisotropy are also considered to demonstrate the generality of the approach. Molecular structures of all complexes considered in this work are shown in Figure \ref{fig:struct}.

All calculations are conducted using a locally extended version of the PySCF quantum chemistry package \cite{Sun2020}. For SA-CASSCF, we choose $3d$ orbitals of the central metal ion as the active space, namely CAS$(ne, 5o)$, with $n$=3, 5, 6, 7, and 8 for V, Mn, Fe, Co and Ni complexes, respectively, and consider all possible spin multiplets formed in the corresponding CAS for state-averaging, e.g. 10 quartets and 40 doublets for $d^7$ Co\suptxt{II}. Scaler relativistic effects are taken into account through spin-free eXact-2-component (SFX2C) Hamiltonian \cite{Dyall2001,Liu2009}. The SOC term is treated by using  spin-orbit mean-field (SOMF) approximation to the Breit-Pauli Hamiltonian \cite{Hess1986, Neese2005}.

\section{\label{sec:results} Results and discussion}

\subsection{Accuracy of DMET+CASSCF/NEVPT2 approach to SIMs}\label{subsec:benchmark}

We conducted all-electron and DMET-based CASSCF and NEVPT2 calculations for the selected set of transition metal complexes and the ZFS parameters obtained from these approaches are summarized in Table \ref{tab:ZFS}. Experimentally measured $D$ values are also collected for comparison. It is important to note that the experimental $D$ values for complexes \textbf{1}-\textbf{10} were derived from variable-temperature susceptibility measurements, while those for \textbf{11}-\textbf{13} were obtained from electron paramagnetic resonance (EPR) spectroscopy. Given the challenges associated with extracting ZFS parameters from experimental data, direct comparison between theoretical and experimental results should be approached with caution. Despite these uncertainties, the overall agreement between theory and experiment is generally satisfactory. Nevertheless, we will focus on the comparison between the results obtained from DMET and all-electron treatment in the following discussion.

For the CASSCF-SO treatment of Co\suptxt{II} complexes, the differences between all-electron and DMET results are generally within 3 $\mathrm{cm^{-1}}$, which is consistent with our earlier findings \cite{Ai2022}, except for complexes \textbf{4} and \textbf{6}, which show a difference of 6.2 cm$^{-1}$ and 4.6 cm$^{-1}$, respectively. A similar level of agreement is observed for other transition metal complexes. Although the $D$ values for these complexes are notably smaller than those of Co\suptxt{II} complexes, DMET still provides quite accurate results that deviate from all-electron ones by less than 0.5 cm$^{-1}$. When NEVPT2 corrections are included, the discrepancy between DMET and all-electron results increases noticeably, with a mean relative error of 8.76\% compared to 4.82\% for DMET+CASSCF. Nevertheless, the differences remain within 10 $\mathrm{cm^{-1}}$, except for one outlier, complex \textbf{4}, which exhibits an error of 15 $\mathrm{cm^{-1}}$. It is also noteworthy that the effect of considering the NEVPT2 correction is strongly system-dependent, and changes the calculated $D$ value by either marginally (e.g. complex \textbf{5}) or significantly (e.g. complex \textbf{6}), which implies that the importance of dynamical correlation as described by NEVPT2 depends on the nature of the metal-ligand bonding in different transition metal complexes.

\begin{table*}
\begin{center}
\caption{ \label{tab:ZFS}
  ZFS parameters $D$(in $\mathrm{cm^{-1}}$) of SIMS by all-electron (AE) and DMET calculations at the CASSCF and NEVPT2 level. In DMET calculations, the impurity is chosen to include all localized orbitals centered on the central transition metal atom. The last two rows show the mean absolute error (MAE) and mean relative error (MAE), in cm$^{-1}$ and percentage, respectively, of DMET results with respect to all-electron ones.
}
\begin{tabular*}{0.7\textwidth}{@{\extracolsep{\fill}}c rr rr  r}
 \toprule
   Complex    & \multicolumn{2}{c}{CASSCF} & \multicolumn{2}{c}{NEVPT2} & Expt.\\
               \cmidrule{2-3} \cmidrule{4-5}
	          &   AE  &  DMET    & AE      & DMET     &                          \\
    \hline
    \textbf{1}& -81.76 & -79.32  & -77.11  & -72.10   & -37.6\cite{Chen2014}\\
    \textbf{2}& 49.64  & 49.80   & 41.77   & 41.69   & 30.0\cite{Darmanovic2019}\\
    \textbf{3}& -90.32 & -90.58  & -78.78  & -83.59  & -57.5\cite{Rigamonti2018}\\
    \textbf{4}& -125.38& -119.14 & -120.18 & -105.51 & -109.0\cite{Novikov2015} \\
    \textbf{5}& -103.84& -101.08 & -102.54 & -92.72  & -161.0\cite{Fataftah2014}\\
    \textbf{6}& -65.15 & -60.52  & -46.11  & -43.22  & -40.5\cite{Jurca2011} \\
    \textbf{7}& -104.95& -102.72 & -113.88 & -106.34 & -115.0\cite{Rechkemmer2016}\\
    \textbf{8}& -90.28 & -88.05  & -77.04  & -72.46  & -69.0\cite{Gupta2023}\\
    \textbf{9}  & 14.17  & 14.23  & 10.94  & 10.79  & 9.9\cite{Eichhofer2014}\\
    \textbf{10} & -15.50 & -15.13 & -11.91 & -10.29 & -13.9\cite{Ruamps2013}\\
    \textbf{11} & -3.16  & -2.78  & -3.46  & -3.10  & -3.46\cite{Duboc2010} \\
    \textbf{12} & 1.97   & 1.48   & 1.74   & 1.13   & 1.80\cite{Ye2010}\\
    \textbf{13} & 0.11   & 0.11   & 0.13   & 0.13   & 0.27\cite{Zein2008}\\
    \hline
        MAE    &         & 1.71  &        & 4.01      & \\
        MRE    &         & 4.8\% &        & 8.8\%   & \\
    \bottomrule
\end{tabular*}
\end{center}
\end{table*}

The relatively larger error in DMET+NEVPT2 than that of DMET+CASSCF can be understood by considering the restriction on the variational space introduced by the DMET construction. In DMET+CASSCF, the CASSCF calculation is carried out in the embedded impurity subspace, and all orbitals, including active ones, are linear combination of impurity+bath orbitals. Compared to CASSCF calculation in the total system, the error of DMET+CASSCF is from the limited variational degree of freedom for orbital rotation. The good agreement between DMET and all-electron CASSCF results indicates that ROHF orbitals used in the DMET construction already agree very well with those from all-electron CASSCF. However, in DMET+NEVPT2 calculation, the perturbation correction is obtained by considering only excitations within the embedded impurity space, and neglecting excitations involving unentangled core and virtual orbitals leads to a noticeably larger error.

As for the computational efficiency, DMET shows significant advantages compared to all-electron treatment, especially in the NEVPT2 calculation. As shown in Table \ref{tab:Nr-orb}, by the DMET construction, the number of orbitals that span the variational space are reduced dramatically to approximately twice the number of basis functions centered on the transition metal ion, regardless of the coordination environment and the size of the total system.
As a result, the CPU time for DMET+CASSCF decreases significantly compared to all-electron CAS. The saving is more dramatic for DMET+NEVPT2, in which the CPU time is reduced by nearly two orders of magnitude compared to all-electron NEVPT2 calculation. We notice that the CPU time for  DMET+NEVPT2 is essentially the same as that for DMET+CASSCF, indicating that the computational overhead caused by the NEVPT2 calculation is negligibly small when using DMET, which can be attributed to the fact the computational cost of NEVPT2 scales as $O(n_{\rm inact}^2 n_{\rm vir}^2)$, and the reduction of orbital numbers by using DMET results in great decreasing of both $n_{\rm inact}$ and $n_{\rm vir}$.

\begin{table*}
	\begin{center}
    \caption{Number of orbitals ($N_{\rm orb}$) and the total CPU time ($T_{\rm CPU}$) used in all-electron (AE) and DMET based CASSCF and NEVPT2 (including SISO) calculations. All the calculations were performed on a 64 core 2.6GHz Intel Xeon Platinum 8358 ($\times$2) platform. It should be noted that PySCF in the default setting uses different memory allocation schemes for small and large systems such that the actual CPU times do not scale monotonically with the system size.}

\begin{tabular*}{0.6\textwidth}{@{\extracolsep{\fill}}c r r  r r r r}
  \toprule
         Complex    & \multicolumn{2}{c}{$N_{\rm orb}$} & \multicolumn{4}{c}{$T_{\rm CPU}$(in minutes)}\\
         \cmidrule{2-3} \cmidrule{4-7}
                    &    &      & \multicolumn{2}{c}{CASSCF} & \multicolumn{2}{c}{NEVPT2} \\
                     \cmidrule{4-5} \cmidrule{6-7}
		            & AE & DMET & AE  &DMET& AE   & DMET\\
                    \hline
        \textbf{1}  & 581 & 93  & 40  & 9  & 278  & 10  \\
		\textbf{2}  & 820 & 102 & 125 & 24 & 1266 & 25 \\
		\textbf{3}  & 887 & 94  & 59  & 27 & 1619 & 27 \\
        \textbf{4}  & 788 & 103 & 100 & 20 & 1028 & 22 \\
        \textbf{5}  & 499 & 96  & 21  & 4  & 147  & 4  \\
        \textbf{6}  & 850 & 100 & 147 & 30 & 1623 & 30 \\
        \textbf{7}  & 829 & 98  & 127 & 23 & 1309 & 23 \\
        \textbf{8}  & 965 & 100 & 70  & 34 & 3069 & 34 \\
\bottomrule
\end{tabular*}
    \label{tab:Nr-orb}
	\end{center}
\end{table*}


\subsection{Choice of impurity}
The results shown in Table \ref{tab:ZFS} indicate that the performance of the DMET+NEVPT2 on the prediction of ZFS parameters is in general quite accurate when compared to its all-electron counterpart, but there are some systems where DMET introduces a relatively larger error. One distinct advantage of DMET is that its accuracy can be improved systematically by including more orbitals into the primary impurity subspace. The results presented above in Section \ref{subsec:benchmark} are obtained by choosing the orbitals centered on transition metal ion as the impurity. We can expect improved accuracy by including orbitals on ligands to the impurity, which will, of course, result in increased computational cost.

The impurity is enlarged based on the following consideration: since the ligands interact with the central atom mainly through $p$-orbitals of coordinating atoms, a natural choice is to include those orbitals into the impurity, so that entanglement between $p$-orbitals with outer orbitals can be described, and their dynamic correlation is included. For further accuracy, we may include all the orbitals centered on the coordinating atoms in the impurity. This rather brute-force choice make up the largest impurity in this study, and we expect that it should be accurate enough for the study of single-ion magnets. In the following discussion, we denote the DMET scheme using only the central metal ion as the impurity as DMET(M), that including $p$-orbitals on coordinating atoms as DMET(ML$_{\text{p}}$), and the scheme including all orbitals on coordinating atoms as DMET(ML).

\begin{figure}[!htb]
    \centering
	\includegraphics[width=0.4\textwidth]{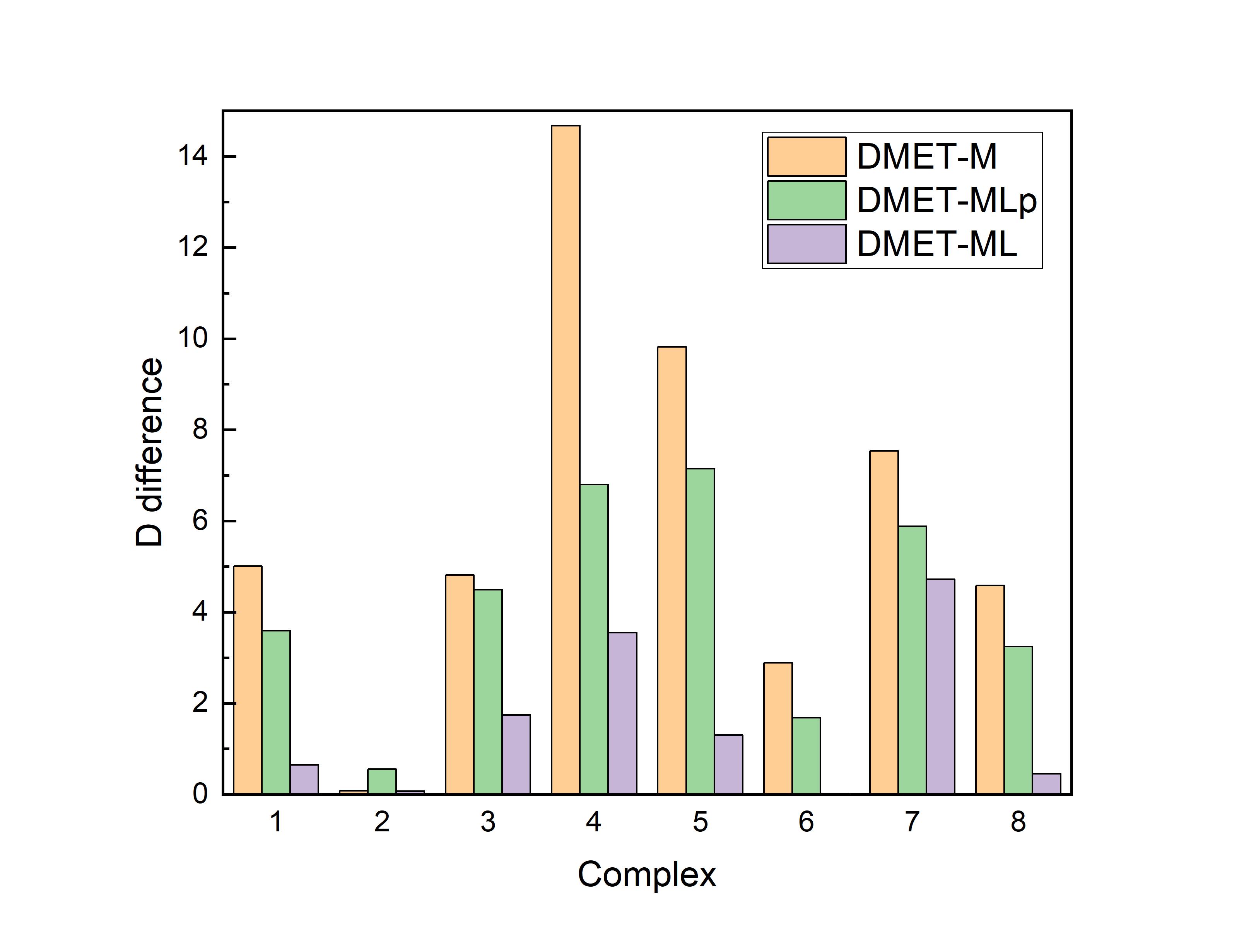}
	\caption{The errors of the calculated ZFS parameters of cobalt complexes using DMET+NEVPT2 compared to all-electron treatment for different choices of the impurity.}
    \label{fig:impurity-choice}
\end{figure}

Figure \ref{fig:impurity-choice} shows a comparison of the errors in the calculated $D$ values obtained by DMET+NEVPT2 with respect to the all-electron treatment for different choices of the impurity for eight cobalt complexes. Including ligand $p$-orbitals has significant effects on the systems that show larger errors in the DMET(M) scheme, i.e. , i.e. complexes \textbf{4} and \textbf{5}, and reduce the error to smaller than 8$\mathrm{cm^{-1}}$, which suggests that in those systems ligand p-orbitals have non-negligible contributions to their magnetic properties. When all the orbitals on coordinating atoms are included in the impurity, the errors are further reduced, generally less than 1-2 cm$^{-1}$ with the only exception of complex \textbf{4}, which has a small but still noticeable error of about 4 cm$^{-1}$. The overall good accuracy indicates that most of the dynamic correlation is accounted for in the embedded impurity space constructed by the DMET(ML) scheme. 

\begin{figure}[!htb]
    \centering
	\includegraphics[width=0.48\textwidth]{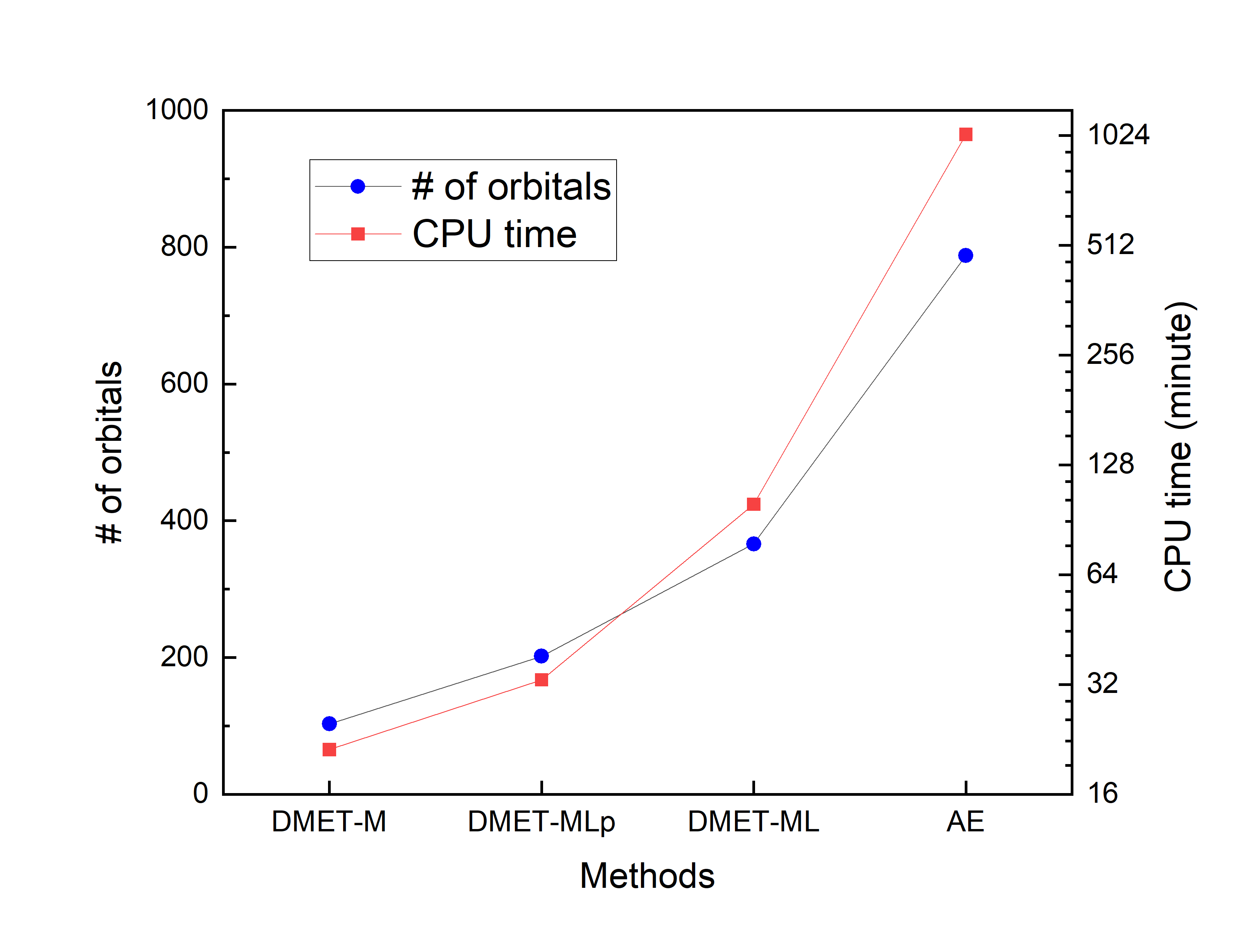}
	\caption{Number of orbitals (the left axis) and the CPU time (right axis) vs the different choice of impurity in complex \textbf{4}.}
    \label{fig:orbitals-time-vs-impurity}
\end{figure}

With the increase of the impurity size, the computation cost grows accordingly, but we can still see a significant reduction of the CPU time when using the DMET(ML) scheme compared to the all-electron treatment. Figure \ref{fig:orbitals-time-vs-impurity} shows the number of orbitals $N_{\rm orb}$ that spans the embedded impurity space and the CPU time using different impurity schemes. One can see that $N_{\rm orb}$ grows fast as we include orbitals on coordinating atoms, since there are 6 coordinating atoms around Co. When all orbitals on coordinating $N$ atoms are included, there are 366 orbitals in the DMET embedded subspace. However, even in that case, using DMET can still reduce the CPU time by one order of magnitude, mainly by the saving in the NEVPT2 calculation. Therefore it is fair to say using DMET can achieve great reduction of computation cost with little loss of accuracy for the NEVPT2 treatment of SIMs.

\begin{table*}
	\begin{center}
    \caption{Pre-SOC excitation energies(in mHa) for complex \textbf{4} calculated by DMET based SA-CASSCF with different choices of the impurity compared to all-electron (AE) results.}
\begin{tabular*}{0.9\textwidth}{@{\extracolsep{\fill}}c r r r r c r r r r}
\toprule
&\multicolumn{4}{c}{CASSCF} &~~~~~& \multicolumn{4}{c}{NEVPT2} \\
 \cmidrule{2-5} \cmidrule{7-10}
 State &  AE   & DMET(M) & DMET(ML\subtxt{p}) & DMET(ML)& & AE & DMET(M) & DMET(ML\subtxt{p}) & DMET(ML) \\
\midrule
 1 & 1.59  &  1.89 &  1.73 &  1.64 & &  2.02 & 2.74  &  2.35 & 2.19  \\
 2 & 31.24 & 31.57 & 31.43 & 31.28 & & 42.54 & 42.94 & 43.21 & 42.57\\
 3 & 31.81 & 32.02 & 31.94 & 31.83 & & 43.83 & 43.84 & 44.31 & 43.79 \\
 4 & 34.84 & 35.14 & 35.02 & 34.89 & & 47.90 & 48.04 & 48.46 & 47.93 \\
 5 & 36.19 & 36.64 & 36.45 & 36.27 & & 49.30 & 49.81 & 50.05 & 49.43 \\
 6 & 43.00 & 43.33 & 43.18 & 43.05 & & 53.79 & 53.97 & 54.35 & 53.84 \\
\bottomrule
\end{tabular*}
	\label{tab:err-energies}
	\end{center}
\end{table*}

\subsection{Investigation of Error Origins in DMET}
In this section, we present some more detailed investigation on possible sources of error in the DMET treatment of $3d$ SIMs, using complex \textbf{4} as an example since it exhibits the most significant error in different DMET schemes. Considering that magnetic anisotropy is determined by both the pre-SOC excitation energies and the SOC matrix elements between the ground state and excited states, we evaluate the accuracy of DMET in calculating these two quantities to identify the sources of error.

For the matrix elements of the SOC Hamiltonian, we found the difference between DMET and all-electron results is only on the order of $\sim 10^{-3}$ mHa, which has a negligible impact on the final prediction of ZFS parameters. Table~\ref{tab:err-energies} shows pre-SOC excitation energies obtained from various computational schemes. Several features are noteworthy. First of all, the excitation energies are generally very small, with the first excitation energy being only about 2 mHa. Such near-degeneracy is characteristic of single-ion magnets, enabling strong SOC perturbations and resulting in large magnetic anisotropy. We can therefore expect that the first excitation contributes most to the zero-field splitting within the ground state multiplet manifold when SOC is considered. At the CASSCF level, the pre-SOC excitation energies from DMET and all-electron treatment differ typically by 0.2-0.3 mHa in the DMET(M) scheme, which decreases to less than 0.05 mHa in the DMET(ML) scheme. Such small differences explain the good agreement in the ZFS parameter by DMET(M) with all-electron calculation(-119.14 vs -125.38 cm$^{-1}$ as shown in Table \ref{tab:ZFS}). When the NEVPT2 correction is considered, the differences between DMET and all-electron treatment increase noticeably when only LOs on Co are treated as the impurity, especially for the first excitation energy, for which the difference is about 0.7 mHa, which results in significantly larger error in the DMET(M) prediction of the ZFS parameter (-105.51 vs -120.18 cm$^{-1}$). Such error can be essentially eliminated by including LOs on the coordinating atoms in the impurity as in DMET(ML), which leads to a very good agreement in the final ZFS parameter, with a difference of less than 4 cm$^{-1}$.


The results shown in Table \ref{tab:ZFS} indicate that there is a systematic overestimation of pre-SOC excitation energies in DMET+CASSCF calculation. This can be understood as followings. In SA-CASSCF, what is variationally minimized is the average energy of all states. Since DMET operates within a restricted variational subspace, the averaged energy is expected to be higher than that from the all-electron treatment. Natural orbital analysis reveals that the ground state of typical Co$^{\text{II}}$ SIMs has nearly integer orbital occupancy and therefore can be well described by the ROHF ground state wavefunction. Given the accuracy of HF-in-HF embedding, we expect that the bath orbitals constructed from the ROHF wavefunction should accurately describe the ground state wavefunction, and the ground state energy from DMET should be close to that of all-electron calculations. Therefore, the higher state-averaged energy arises from the increased energy of the excited states. Physically, the overestimation of excitation energies indicates that the bath orbitals constructed from the ground state wavefunction are insufficient to accurately describe the excited states, which, however, can be systematically improved by extending the impurity size. The results presented above indicate that as far as magnetic anisotropy properties of $3d$ SIMs are concerned, choosing LOs centered on the central transition metal ion as the impurity is adequate in most cases, which, if not accurate enough (e.g. for complex \textbf{4}), can be improved by extending the impurity to include direct coordinating atoms.


\section{Concluding Remarks}\label{sec:summary}

To summarize, we have further developed the combined DMET and CASSCF-SO approach to single-ion magnets by incorporating dynamic correlation in the embedded impurity space using NEVPT2, and we found the computational efficiency of NEVPT2 is significantly enhanced in the DMET framework, primarily due to its quartic scaling with respect to the system size. The accuracy achieved is satisfactory for most systems studied and can be systematically improved by expanding the impurity space to include more local orbitals on coordinating atoms. A DMET scheme that treats the central metal ion and the $p$-orbitals of coordinating atoms as the impurity provides satisfactory results for all systems at a similar computational cost. Higher accuracy can be achieved by including all orbitals of the coordinating atoms in the impurity, which reproduces ZFS parameters within an error margin of $\mathrm{3\,cm^{-1}}$, while reducing the CPU time by a factor of 10 for systems with approximately 800 orbitals when compared to the all-electron treatment.

We close the paper by making a few general remarks regarding the limitations of our current approach and the possibilities for further improvement. CASSCF considers only static correlations that are confined within the active space spanned by a small set of active orbitals, which can be accurately captured by DMET if the embedded impurity (impurity plus bath) subspace fully encompasses the active space. However, dynamic correlation described by NEVPT2 is more non-local, involving all orbitals in the system, making it more challenging for embedding methods. Our findings in this work demonstrate that by a proper choice of the primary impurity, the embedded impurity space obtained by the DMET construction can grasp the most important part of dynamical correlation that is relevant to accurate description of magnetic anisotropy in typical transition metal SIMs. On the other hand, such treatment of dynamical correlation may well be inadequate for properties that depend on electronic excitations of higher energy. For those properties, one may obtain more accurate description by further expanding the impurity size, which would increase computational cost accordingly. To take the full advantage of the DMET construction, it is desirable to optimize the impurity-expanding strategy instead of the brute-force scheme used in this work. Another possibility, which is conceptually more intriguing but may require substantial methodological development efforts, is to directly treat dynamical correlation involving the interaction between the embedded impurity space and the outer space spanned by unentangled core and virtual environmental orbitals, by either adopting techniques similar to multi-reference configuration interaction (MRCI) \cite{Lischka2018} or exploiting the renormalization or down-folding strategy that is widely used in theoretical condensed matter physics \cite{Jiang2015}.


\begin{acknowledgements}
This work is supported supported by the National Natural Science Foundation of China (Project Number 12234001). We acknowledge the High-performance Computing Platform of Peking University for providing the computational facility.
\end{acknowledgements}

\begin{flushleft} \textbf{DATA AVAILABILITY}\end{flushleft}
The data that support the findings of this study are available from the corresponding author upon request.

%

\end{document}